\documentclass[prl,twocolumn,showpacs,preprintnumbers,amsmath,amssymb,floatfix]{revtex4}
\usepackage{hyperref}
\usepackage{epsfig}

\begin{document}

\title{Controlling Microscopic Friction through Mechanical Oscillations}

\author{R. Guerra$^1$, A. Vanossi$^1$, and M. Urbakh$^{2}$}
\affiliation{$^1$CNR-INFM National Research Center S3 and
Department of Physics, \\University of Modena and Reggio Emilia,
Via Campi 213/A, 41100 Modena, Italy \\$^2$ School of Chemistry,
Tel Aviv University, 69978 Tel Aviv, Israel }

\date{\today}

\begin{abstract}

We study in detail the recent suggestions by Tshiprut {\it et al.}
[Phys. Rev. Lett. {\bf 95}, 016101 (2005)] to tune tribological
properties at the nanoscale by subjecting a substrate to periodic
mechanical oscillations. We show that both in stick-slip and
sliding regimes of motion friction can be tuned and reduced by
controlling the frequency and amplitude of the imposed substrate
lateral excitations. We demonstrate that the mechanisms of
oscillation-induced reduction of friction are different for
stick-slip and sliding dynamics. In the first regime the effect
results from a giant enhancement of surface diffusion, while in
the second regime it is due to the interplay between washboard and
oscillation frequencies that leads to the occurrence of parametric
resonances. Moreover we show that for particular set of parameters
it is possible to sustain the motion with the only oscillations.

\end{abstract}


\maketitle

\section{Introduction}

In the emerging field of nanoscale science, major efforts have
been devoted to form a deep, predictive understanding of the
mechanisms involved in fundamental tribological processes,
envisaging alternative innovative solutions to control friction
\cite{Urbakh}. The ability to manipulate frictional forces is
extremely important for many technological applications. One may
wish to reduce or enhance friction, eliminate the high-dissipative
chaotic and stick-slip regimes of motion, and instead, to achieve
smooth sliding. Early stages of motion and stopping processes in
micromechanical devices and computer disk drives, which exhibit
chaotic stick-slip, pose, for example, a real problem. The
difficulties in realizing an efficient control of friction are
related to the complexity of the task, namely dealing with systems
with many degrees of freedom under a strict size interfacial
confinement. The achievements of nanotechnology have opened new
perspectives and new basic scientific questions in this direction,
where novel local probes give access to the study of friction at
the atomic level.

A novel approach for tuning frictional response, which has
recently attracted considerable interest \cite{Rozman, Gao,
Socoliuc, Heuberger, Zaloj, Cochard, Igarashi}, is the mechanical
control of a system, via externally imposed normal or lateral
vibrations of small amplitude and energy to the sliding interface.
Manipulations by mechanical excitations, when applied at suitable
frequencies and amplitudes, can drive particles, which are close
to the interface, out of their potential energy minima, thus
increasing considerably surface diffusion and mobility, and
reducing friction. Through the dynamical stabilization of
desirable regimes of motion (e.g., smooth sliding), which would be
otherwise unstable, substrate mechanical oscillations can reduce
high dissipative sliding behaviors (stick-slip).

Since the idea is not to change the physical properties of mating
interfaces, flexibility and accessibility are main relevant
features of this approach. Controlling frictional forces has been
traditionally approached by chemical means \cite{Dudko,Mosey},
usually by supplementing base lubricants with friction modifying
additives. However, standard lubrication techniques provide
prescribed tribological properties which fit a certain range of
parameters of the sliding objects (such as the load) and the
external forcing, and are expected to be less effective at the
nanoscales. Here, on the contrary, frictional properties can be
tuned continuously by the frequency and the amplitude of the
out-of-plane or in-plane vibrations of the sliding surfaces. The
predicted effects should be amenable to atomic force microscopy
(AFM) tests \cite{Riedo,MUR2003} or applying ultrasound to the
sample \cite{Ge,Dinelli}, or even in studies of contact mechanics
of a probe interacting with oscillating quartz crystal
microbalance surfaces \cite{Borovsky,Berg,Lubben}.

In this paper, we extend the recent results \cite{Tshiprut}
concerning the ability to modify tribological properties at the
nanoscale by subjecting a substrate to lateral mechanical
oscillations. The investigation is limited to a one-dimensional
(1D) Tomlinson-like model. We show that microscopic friction can
be suitably tuned and reduced by controlling the resonant
frequencies and amplitudes of the imposed substrate lateral
excitations. Depending on the model parameters, the numerical
simulations reveal the details of the observed different regimes
of motion (smooth sliding, stick-slip, inverted stick-slip) and of
the occurring transitions among them. Moreover we show that for
particular sets of parameters values the motion can be sustained
just with the oscillations.

\section{Model and numerical method}

From a theoretical point of view, it has been frequently shown
that, despite the simplified level of description,
phenomenological models of friction have revealed the ability of
mimicking the main features of the tribological behavior observed
both in nanoscale experiments or in more complex MD simulation
frameworks \cite{Vanossi}.

The {\it Prandtl-Tomlinson model} is, probably, the most widely
used in interpretation of tribological experiments due to its
simplicity and its ability of accounting for the main physical
features of atomic-scale friction (see, e.g.,
\cite{MUR2003,R2000}). In this mathematical description, a
particle of mass $m$ is coupled, via a harmonic spring of
stiffness $K$, to a moving stage which slides at constant velocity
$v_{G}$. Moreover, the particle experiences a periodic
(sinusoidal) coupling $V(x)= U \sin \left( 2 \pi x / b \right)$ to
the substrate, where $x$ is the current position of the particle.
Introducing the viscous damping coefficient $\gamma$, which
results from the energy dissipation to phonons or other
excitations in the substrate, the equation of motion for the particle
becomes
\begin{equation}
m \ddot x + m \gamma \dot x = - \frac{{dV(x)}}{{dx}} - K (x -
v_{G} t). \label{Tomlin}
\end{equation}
If the pulling spring is stiff enough (i.e., $K$ is larger than
the effective substrate spring constant $K_0 = max \left[ V''(x)
\right]$), then there is always a unique equilibrium position
$x_{\rm eq} \approx v_{G} t$ for the particle. Consequently, at
small driving velocity, the friction will be linear in $v_{G}$.
 Things become more interesting once $K_0$ exceeds $K$.
In this situation, the potential energy landscape experienced by
the particle is characterized, at certain instances of time, by
more than one stable position. The time dependence of the combined
spring and substrate potential reveals that mechanically stable
positions disappear at certain instances in time due to the motion
of the spring. The system is locked in one of the minima of the
potential energy landscape until the increasing elastic stress of
the spring allows overcoming the barrier. After that, the slider
is accelerated. The potential energy of the elastic stress is
converted into the kinetic energy of the slip event and,
eventually, dissipated into heat via the damping $\gamma$-term.
Then, the particle rapidly drops to the next nearest metastable
minimum, where it locks again. As a consequence, for sufficiently
small driving velocity, the dissipated energy per sliding distance
is rather independent of $v_{G}$, in agreement with many
experimental results. Obviously, this simple mechanical
description can provide only qualitative interpretations of the
underlying tribological processes. The Tomlinson model is
described in detail in many surveys, e.g., in Refs.\
\cite{MUR2003,R2000}, with a long list of applications to concrete
physical systems (in particular to those concerning tip-based
devices).

\begin{figure}
\epsfig{file=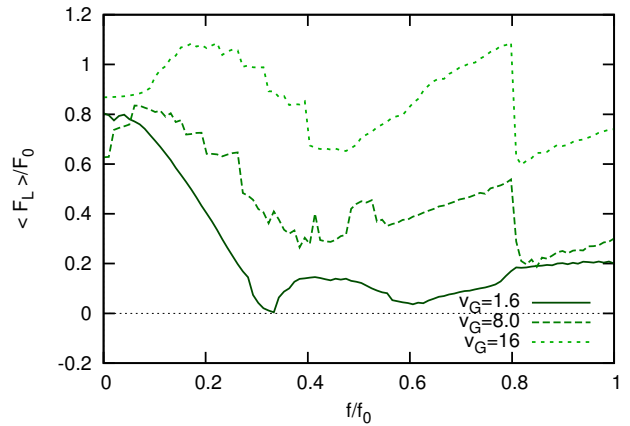,width=6.0cm,angle=-90,clip=}
\caption{\label{Fig1a}  (Color online) Averaged friction force
dependence on the oscillation frequency externally imposed to the
substrate. The behavior is shown for three different values of the
driving stage velocity $v_{G}$: 1.6 (solid), 8.0 (dashed), 16.0
(dotted). The lateral force $F_L = K(x- x_G )$ is normalized to
the static friction force $F_0 = max[V'(x)]$, and the oscillation
frequency $f$ to the system characteristic frequency $f_0 =
\frac{1}{b} \sqrt{\frac{U}{m}}$.}
\end{figure}

In this work we consider a Tomlinson-like model as that recently
introduced in ref. \cite{Tshiprut}. In particular, in order to
study the effect of lateral vibrations on 1D diffusion and
friction, the model describes the diffusion and driven dynamics of
an AFM tip (a single mass driven by an external guide) subject to
a substrate, which oscillates in the lateral direction with
amplitude $A$ and frequency $f$. The potential of such oscillating
substrate becomes
\begin{equation}
V(x,t) = U \sin \left[ \frac{2\pi}{b} \left( x+A\sin(2\pi f t)\right) \right].
\label{ACsubstrate}
\end{equation}

The motion of the tip in lateral direction is governed by the
equation
\begin{equation}
m \ddot{x} = F_{sub} \left(x, t\right) - K \left( x - x_{G}
\right) + \Gamma \left(t\right) - m \gamma (\dot{x}-v_{sub}),
\label{our_model}
\end{equation}
where $F_{sub}$ is the force due to the substrate potential
$V(x,t)$, and $x_G = v_G t$; $v_{sub} = -2\pi fA\sin(2\pi f t)$
represents the velocity at which the substrate oscillates. In
particular, the last two terms take into account the thermal and
dissipation effects in the framework of the Langevin dynamics
\cite{Risken}. The random force $\Gamma(t)$ and the viscous
damping term $\gamma$ are related by the fluctuation-dissipation
theorem:
\begin{equation}
\langle\Gamma(t) \Gamma(t')\rangle =2 \gamma mk_{B}T \delta (t -
t') \label{eq:fluct-diss}
\end{equation}
with $k_{B}$ and $T$ denoting the Boltzmann constant and the
temperature respectively. These two added terms can be thought of
as representing degrees of freedom inherent in a real physical
system which are not explicitly included in the adopted simplified
mathematical description. In particular, we assume that the
damping term in Eq. (\ref{our_model}) is determined by the energy
dissipation to phonons and electron-hole excitations in the
substrate. As a result, this term is proportional to the relative
velocity of the driven tip and the substrate. Under these
conditions, the substrate vibrations cause a time-periodic (ac)
force acting on the tip, $F_{ac} = mA(2\pi f)^2\sin(2\pi ft)$.
This force presents the effect of inertia.

In the strong (weak) dissipative regime, when $\gamma$ is much
larger (smaller) than the characteristic vibrational frequencies
of the system, the motion is overdamped (underdamped). The
numerical integration is carried out using the well-known velocity
Verlet algorithm. For every value of the parameters and after
reaching the steady state, the system characteristics of relevant
physical interest are measured. Natural units used to characterize
the model for length, time and energy are $\AA$, ${\mu}s$, and
$meV$. If not stated differently, $m=1$, $U=100$, $b=1$, $K=12.63$, 
$T=0$, and $\gamma=32$ define our default parameter values. For
parameters close to these default values, the system can be
considered to behave as overdamped. Under these parameter ranges,
the predicted effects, as for surface diffusion resonance
frequencies, should be amenable, e.g., to atomic force microscopy
(AFM) tests using shear modulation mode, for which we typically
have $m \approx 10^{-12} \div 10^{-14}$ kg, $U \simeq 0.25$ eV,
and $b \simeq 0.4$ nm. For these values of system parameters, the
characteristic frequency turns out to be $f_0 = (1/b)
\sqrt{U/m} \approx 1-10$ ${\mu}$s$^{-1}$ and velocity unit is
${\AA}/{\mu}s$.

\section{Results and discussion}

In order to investigate the effectiveness of lateral mechanical
oscillations and their ability of improving nanoscale tribological
properties, we have performed numerical simulations in different
ranges of model parameters where the ``natural'' system (without
oscillations) may exhibit both high dissipative stick-slip
dynamics and smooth sliding behaviors.

At a fixed value of the amplitude $A(=1)$, Fig. \ref{Fig1a}. shows
the influence of the substrate oscillation frequency $f$ on the
time-averaged friction force $<F_L>=<K(x- x_G)>$, for three
distinct values of the driving velocity $v_G$, corresponding
respectively to stick-slip dynamics for $v_G=1.6$ and to smooth
sliding for $v_G=8$ and $v_G=16$. The frequency $f$ is varied
adiabatically and, for each of its value, $F_L$ is averaged over a
sufficiently long simulation time after reaching the dynamical
steady state. Depending on the external stage velocity, the
results presented in Fig. \ref{Fig1a} exhibit two types of
resonance minima of friction which are determined by different
physical mechanisms. As we will show shortly the low frequency
resonance, which is most pronounced for the low driving velocity
(e.g., $v_G = 1.6$), results from the giant enhancement of surface
diffusion induced by the oscillations of the substrate, as
recently discussed in details in ref. \cite{Tshiprut}. On the
contrary, for higher driving velocities (e.g., $v_G = 8.0$;
$16.0$), sharp resonance minima in the average lateral force
$<F_L>$ are caused by the interplay between the system washboard
frequency $v_G/b$ and the oscillation frequency $f$.

An understanding of the effects of lateral substrate vibrations on
the frictional properties of the system at low/moderate driving
velocities, comes from the study of the diffusion coefficient for
the undriven tip as a function of the oscillation frequency
\cite{Tshiprut}. For low external frequencies $f$, the tip apex is
able to follow the slow motion of the substrate, performing just
small vibrations around the potential minima. Under this
condition, if the thermal energy is sufficiently smaller than the
height of the potential barrier, the probability to escape from
the potential well is exponentially small. For the optimal
matching of the oscillation frequency $f$ and the amplitude $A$,
the diffusivity $D$ may be considerably enhanced \cite{Hanggi}. In
this case, the tip approaches the maximum of the substrate
potential at the end of half cycle of the surface oscillations,
where the driving force acting on the tip diminishes. Then, even a
weak thermal noise splits the ensemble of tips into two parts that
relax to the neighboring minima of the surface potential, and the
resonance enhancement of diffusion is observed. A further increase
of the frequency leads again to localized dynamics; in contrast to
the case of low frequencies here the tip overcomes the potential
barriers and oscillates between neighboring minima of the surface
potential. The resonance enhancement of surface diffusion leads to
a corresponding reduction of the mean friction force at $f \simeq
3.32$ which is most pronounced for low driving velocities (see
Fig. \ref{Fig1a} for $v_G=1.6$).
\begin{figure}
\epsfig{file=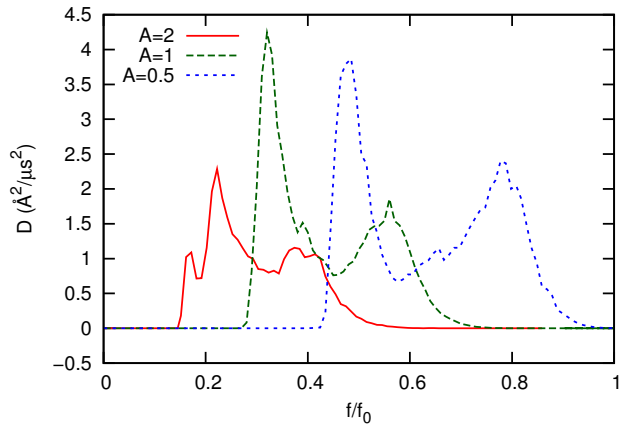,width=6.0cm,angle=-90,clip=}
\caption{\label{Fig2a} (Color online) Diffusion $D$ of the tip as
a function of the normalized oscillation frequency $f/f_0$, $f_0 =
\frac{1}{b} \sqrt{\frac{U}{m}}$, at finite temperature
($k_{B}T/U=0.01$), for three different values of the oscillation
amplitude $A$.}
\end{figure}
We found that the curves of Fig. \ref{Fig2a} calculated for three
oscillation amplitudes $A$ are not changed qualitatively with
increase of temperature, the diffusion resonance peaks are still
visible for higher temperatures but greatly broaden and damped.
The driving force acting on the tip is proportional to the
amplitude of oscillations and, as a result, we observe that the
resonance frequency decreases with $A$.

\begin{figure}
\epsfig{file=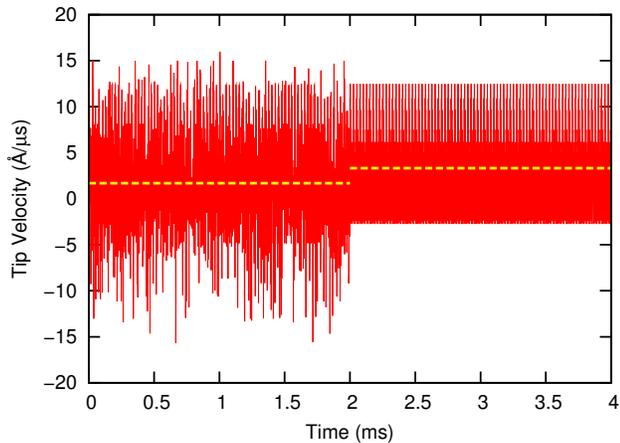,width=6.0cm,angle=-90,clip=}
\caption{\label{Fig3a}  (Color online) During the first $2$ ms the
tip is driven at $v_G=1.6$ and then detached from the guide. The
substrate oscillates uninterruptedly over the whole simulation
with $f=3.32$ and $A=1.44$. Clearly, the tip motion is completely
carried by the substrate vibrations, with an average velocity
highlighted by the dashed line.}
\end{figure}

By systematically scanning the phase space of the model
parameters, it is possible to tune the characteristics of the
external oscillating excitations, in terms of suitable choices of
amplitude and frequency, to achieve very favorable sliding
conditions, for which friction almost vanishes or even $F_L \leq
0$ on average. For a negative value of the lateral friction force,
the driving guide, moving at speed $v_G$, does not pull the tip
apex anymore, but it is instead pushed forward by the tip itself.
An example of such a sliding situation is shown in Fig.
\ref{Fig3a}, where, once the external guide moving at $v_G=1.6$ is
detached from the tip, the direct tip motion is fully sustained by
the only sinusoidal oscillations of the substrate, at the
diffusion resonance $f=3.32$. Due to highly nonlinear and strongly
fluctuating dynamics, the determination of desirable set of model
parameters achieving this ``negative'' friction regime is not
always a trivial task.

For higher driving velocities, as those shown in Fig. \ref{Fig1a}
(i.e., $v_G = 8.0; 16.0$), the mechanism of reduction of friction
is essentially different, stemming from the interplay between the
system washboard frequency $v_G/b$ and the oscillation frequency
$f$. This resonant behavior of the driven tip is clearly displayed
by sudden and sharp drops of $<F_L>$ at specific oscillation
frequencies $f_R$, which are given by the formula $f_R = v_G/Nb$,
with $N$ integer.

\begin{figure}
\epsfig{file=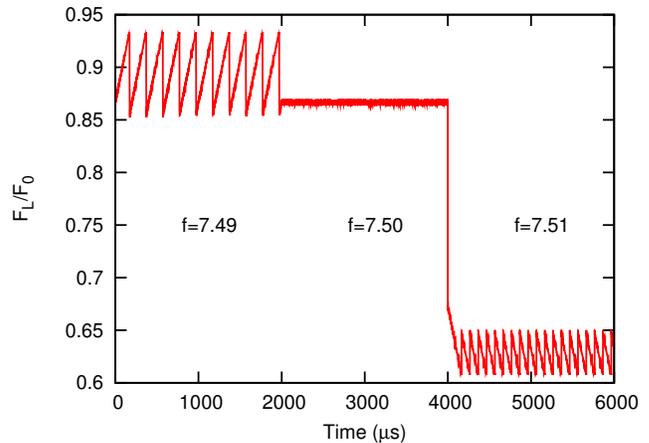,width=6.0cm,angle=-90,clip=}
\caption{\label{Fig4a}  (Color online) $F_L/F_0$ time-dependence
passing through a resonance frequency. At $A=1.44$ and $v_G=15$,
the oscillation frequency is increased from $f=7.49$ ($0-2000$ $\mu$s)
to $f=7.50=v_G/2$ ($2000-4000$ $\mu$s), up to $f=7.51$ ($4000-6000$
$\mu$s). The lateral force oscillations are almost completely washed
out at the resonance frequency. The system shows, across the
resonance, a transition from a stick-slip-like to an inverted
stick-slip-like regime.}
\end{figure}

For an in-depth investigation of these features, we have analyzed
the detailed dynamics during the transition across a resonance
frequency $f_R=v_G/2b$ for the case of an external guide velocity
$v_G=15$ (see Fig. \ref{Fig4a}). Due to a dynamical combination of
fast sliding and substrate oscillations, for frequencies $f
\lesssim f_R$ the lateral force time series shows an asymmetric
sawtooth shape resembling that of the usual stick-slip but with
very tiny amplitude; as soon as $f = f_R$ the lateral force stops
to oscillate, meaning that the guide-tip separation remains
constant. Finally, when $f \gtrsim f_R$ the synchronization is
lost and $F_L$ starts again oscillating with doubled frequency,
but showing an inverted stick-slip-like behavior. The resonance
marks in general the transition from a negative to a positive
phase difference between the characteristic washboard frequency
$v_G / b$ and the oscillation frequency $f$, which is
characteristic for parametric resonances. This effect can lead to
significant reduction of friction in the sliding regime of motion,
as shown by sharp drops in average and instantaneous lateral
forces presented in Figs. \ref{Fig1a} and \ref{Fig4a}. It should
be noted that this mechanism of reduction of friction in the
sliding state is very different for that discussed above for the
stick-slip regime.

\begin{figure}
\epsfig{file=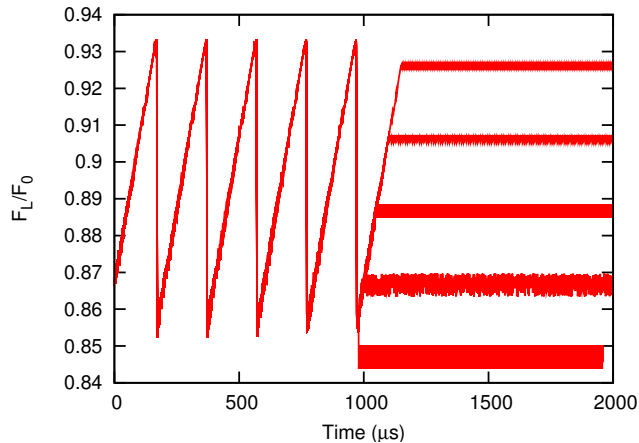,width=6.0cm,angle=-90,clip=}
\caption{\label{Fig5a} (Color online) Lateral force vs. time in
approaching the resonance frequency $f = 7.50 = v_G/2$. Switching
to the resonance at different times may result in a discrete set
of distinct and evenly separated values of $F_L$ due to different
initial conditions. }
\end{figure}

During the adiabatic increasing procedure of $f$, it has been
noted that switching to the resonance $f_R$ at different times may
result in distinct values of lateral friction force because of the
different initial conditions. Due to commensuration between the
washboard $v_G / b$ and the frequency $f$ at the resonance $f_R$,
this behavior shows in Fig. \ref{Fig5a} the occurrence of only a
discrete set of allowed values for $<F_L>$; the evenly separation
between these states is determined by the dimensionless constant $K
b/F_0$.

\section{Conclusions}

In summary, the model demonstrates the effectiveness of imposed
lateral excitations for the reduction and control of friction at
the nanoscale under different regimes of motion (from stick-slip
to smooth sliding) of a mating contact. For an optimal matching of
the oscillation frequency and amplitude, this mechanical
manipulation may drive, for example, an interfacial system out of
its potential energy minima, thus increasing considerably surface
diffusion and mobility, and reducing friction. Unfavorable high
dissipative stick-slip motions can be suppressed in favor of the
dynamical stabilization of desirable smooth-sliding regimes.

It has been shown that physical mechanisms of the oscillation
induced reduction of friction are different for stick-slip and
sliding regimes of motion. For stick-slip dynamics, at low driving
speed, giant enhancement of surface diffusion takes place. In this
case, the observed tribological response may turn to be quite
sensitive to thermal effects, able to spoil the sharp resonant
nature of this phenomenology. Differently, for smooth sliding
regimes at high driving velocity, it is the interplay between
washboard and oscillation frequencies that determines the
occurrence of favorable parametric resonances. A very careful
tuning of the model parameters has been proved to lead to the
possibility of a directed motion sustained by the sinusoidal
substrate oscillations only.

\section*{Acknowledgments}
We are grateful to Alexander E. Filippov for invaluable
discussions. This research was partially supported by PRRIITT
(Regione Emilia Romagna), Net-Lab ``Surfaces \& Coatings for
Advanced Mechanics and Nanomechanics'' (SUP\&RMAN). A.V. thanks
the School of Chemistry at Tel Aviv University for the kind
hospitality and the ESF ``Nanotribology'' Program for financial
support. M.U. acknowledges the support by the Israel Science
Foundation (Grant No 1116/05).


\end{document}